\documentclass{svproc}
\usepackage{url}
\usepackage{graphicx}
\graphicspath{ {figures/}}
\usepackage{booktabs}
\usepackage{url}

\usepackage{balance}
\usepackage{breakurl}
\usepackage{amsmath}

\begin{document}
\mainmatter              
\title{Mind The Gap: Can Air-Gaps Keep Your Private Data Secure?}

\titlerunning{Mind The Gap: Can Air-Gaps Keep Your Private Data Secure?}  
%
\author{Mordechai Guri}
\authorrunning{Mordechai Guri} 
%

%
\institute{Ben-Gurion University of the Negev, Israel\\ Department of Software and Information Systems Engineering \\  Air-Gap Research Lab
 \\ Email: \email{gurim@post.bgu.ac.il}\\ home page:
\texttt{http://www.covertchannels.com}}

\maketitle              

\begin{abstract}
Personal data has become one of the most valuable assets and lucrative targets for attackers in the modern digital world. This includes personal identification information (PII), medical records, legal information, biometric data, and private communications.

To protect it from hackers, 'air-gap' measures might be employed. This protective strategy keeps sensitive data in networks entirely isolated (physically and logically) from the Internet. Creating a physical 'air gap' between internal networks and the outside world safeguards sensitive data from theft and online threats. Air-gap networks are relevant today to governmental organizations, healthcare industries, finance sectors, intellectual property and legal firms, and others.

In this paper, we dive deep into air-gap security in light of modern cyberattacks and data privacy. Despite this level of protection, publicized incidents from the last decade show that even air-gap networks are not immune to breaches. Motivated and capable adversaries can use sophisticated attack vectors to penetrate the air-gapped networks, leaking sensitive data outward. We focus on different aspects of air gap security. First, we overview cyber incidents that target air-gap networks, including infamous ones such as Agent.btz. Second, we introduce the adversarial attack model and different attack vectors attackers may use to compromise air-gap networks. Third, we present the techniques attackers can apply to leak data out of air-gap networks and introduce more innovative ones based on our recent research. We demonstrate that despite the disconnection from the Internet, attackers can exploit special types of covert channels to exfiltrate confidential data. We present the different types of covert channels, including electromagnetic, electrical, optical, thermal, and acoustic techniques. We show how attackers can use them stealthily. Our paper shows how advanced persistent threats (APTs) can leak private and confidential information using these techniques, even when protected behind air gaps. Finally, we propose the necessary countermeasures to protect the data, both defensive and preventive.
\keywords{air-gap, private information, personal information, confidential data, data leakage, networks, exfiltration, covert channel,  protection, cyberattacks}
\end{abstract}

\section{Introduction}
As the world becomes more digitized, information becomes a precious source of income for criminals and a highly lucrative target for them. Consequently, in the last decade, more cyberattacks have been aimed at data theft, information stealing, and espionage. These attacks range from malware deployed on organizational networks to spy apps targeting executives' computers. To deal with that, organizations protect their IT networks with many defensive measures. These include firewalls, security appliances, and intrusion prevention systems at the network level. In addition, antivirus software, anti-exploitation, and storage encryption at the host level. However, regardless of the level of security used, as long as organizational networks are connected to the Internet, they are exposed to a wide range of online threats and cyberattacks. According to the reports, thousands of publicly disclosed data breaches occurred in 2021, with approximately 22 billion records being leaked \cite{Over22bi79:online}. Notably, several top-tier companies have been affected by data breach attacks in recent years, including Microsoft, Amazon, Apple, and Google \cite{TheTop9R26:online}.

\subsection{Air-Gap Data Protection}
There are many cases when data must be kept strictly protected and secured to eliminate the danger of information leakage.
For example, under current regulations, certain data must be strictly secure. GDPR (Global Data Protection Regulation) sets strict rules on how personal data is collected, processed, and stored by companies and institutions \cite{voigt2017eu}. Other data types subject to such requirements may include financial data, classified governmental and military information, etc. 
When sensitive and confidential information is maintained, the organization may keep it separate from the Internet. With this setup, known as an 'air gap,' the local network is isolated, with no connection between the local network and the Internet. This isolation level eliminates all online threats and cyberattacks since hackers have no direct access to the valuable network.
The air-gap policy may be used by governmental offices, finance institutes, defense and military, public sectors, global businesses, and other industries \cite{AirGappe40:online}. In the context of data security and privacy, information that might be protected in air-gap networks may include; (1) personal identification information, (2) financial information, (3) medical information and health records, (4) legal information (e.g., court documents, legal agreements, confidential legal advice), (5) biometric data (e.g., fingerprints, facial recognition data).


\subsection{Air-gap Attack Model}
Although air-gapped networks are considered safer, they are not immune to breaches. More than 17 known malicious frameworks appear to target data on air-gapped networks reported in the media over the years \cite{dorais2021jumping}. In the past, several reported air-gap breach incidents involved compromising air-gapped networks. In the case of Agent.btz, a classified network of the United States military was infected with an APT worm \cite{Guri:2018:BAM:3200906.3177230}. Other air-gap hacking cases reported in the past include Stuxnet \cite{kushner2013real}, USBFerry and ProjectSauron \cite{dorais2021jumping}. Attackers may use sophisticated techniques to breach air-gap isolation, such as supply chain attacks, software contamination (E.g., SolarWinds supply chain attack \cite{wolff2021navigating}), and insider personnel. With these methods, attackers can evade intrusion prevention and detection (IDS/IPS) systems and gain a foothold in the network.

\subsection{Data Exfiltration via Physical Covert Channels}
After breaching the air-gapped network, the malware collects data from the compromised network and storage. Although breaching air-gapped networks has been demonstrated to be feasible, information exfiltration from disconnected networks to the Internet is a more challenging problem. This is because there is no network connectivity, wired or wireless, that the attacker can use to leak data outwards. The research domain of special covert communication techniques that work from isolated, air-gapped networks is known as air-gapped covert channels. By employing these techniques, the attacker can modulate information and transfer it to a third party. The use of \textit{electromagnetic} mediums to leak data has been explored for many years. In this approach, malicious code uses electromagnetic leakage from laptops, PC, and servers to leak data. Other techniques include using \textit{acoustic} waves, \textit{magnetic} fields, \textit{thermal} emissions, and \textit{vibrations} for data exfiltration \cite{guri2022air}. For example, an attacker can encode private biometric information (e.g., fingerprints) and leak it over an inaudible ultrasonic sound several meters away to a nearby smartphone in seconds.

\begin{figure}
	
	\centering
	
	\includegraphics[width=\linewidth]{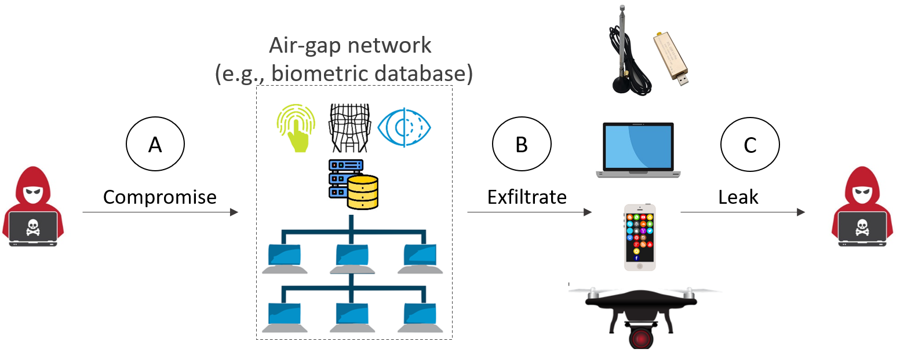}
	
	\caption{The chain of attack. The attacker compromises the air-gapped network (a biometric database in this case) (A), collects confidential information, and exfiltrate it via covert channels (e.g., electromagnetic, acoustic, or optical) (B), which are then forwarded to the remote attacker (C).}
	
	\label{fig:three}
	
\end{figure}

The chain of attack is illustrated in Figure \ref{fig:three}; The attacker compromises the air-gapped network (a biometric database in this case) (A), collects confidential information, and exfiltrates it via covert channels (e.g., electromagnetic, acoustic, or optical) (B), which are then forwarded to the remote attacker (C)).

\section{Air-gap as Data Security Measure}
Air-gapping, or an air-gap (or air-wall), is a security measure to isolate a computer or network from external networks, such as the World Wide Web or other connected systems. It involves physically or logically disconnecting a computer or network from any external or untrusted network to create a physical or digital gap, preventing direct communication or data transfer between the isolated system and the outside world.

The term "air gap" originates from the idea that isolation creates a gap through which data or information cannot pass without physical or intentional interaction. This approach might be used in high-security environments, such as government facilities, critical infrastructure systems, and secure research laboratories, where protecting sensitive or classified information is paramount.

An air gap shields the isolated system from various cybersecurity threats, including remote hacking, malware infections, and unauthorized data exfiltration. Since the system is disconnected from external networks, it is more challenging for attackers to gain access or transmit data into or out of the isolated environment.

\subsection{Air-gap and Usability}
While air gapping provides a high level of security, it can also present challenges regarding convenience, usability, and data transfer between systems. Transferring data to and from an air-gapped system typically requires manual methods, such as using removable media (e.g., USB drives) or dedicated communication channels. Security measures must be taken to ensure the integrity of the data and prevent the potential introduction of malware or other security risks during the transfer process, for examples use IDS, firewalls, and data diodes \cite{jeon2016study}\cite{bostanunidirectional}.

\subsection{Protecting Sensitive and Private Data}
Air-gapping is commonly used in various industries and sectors that handle sensitive or classified information and have stringent security requirements, as listed in Table \ref{tab:inf}. No specific regulations explicitly require air-gapped networks. However, certain industries and sectors have regulatory frameworks that mandate stringent security measures to protect sensitive, private, or classified information. These regulations may indirectly influence air-gapping implementation or require similar security controls to ensure the confidentiality, integrity, and availability of data. Government entities and defense organizations often have strict regulations and standards for protecting information \cite{PolicyDo14:online}\cite{bostanunidirectional}. These include guidelines and requirements for secure communication, data storage, and information sharing, and also may necessitate air-gapped networks or similar security measures \cite{Security19:online}. From a technical perspective, isolation is also known as red-black separation \cite{bostanunidirectional}\cite{101994pd50:online} which prevents the unauthorized transfer of information between different security domains or classification levels. Red-black separation involves the physical and logical separation of networks, systems, or data repositories \cite{bostanunidirectional}. This is to maintain strict control over information flow and mitigate the risk of unauthorized access or leakage. 

In the healthcare sector, regulations like the Health Insurance Portability and Accountability Act (HIPAA) of the United States \cite{annas2003hipaa} and the General Data Protection Regulation (GDPR) of the European Union \cite{voigt2017eu} impose requirements for protecting patient data and electronic health records. While these regulations do not explicitly mandate air gapping, they emphasize the importance of implementing appropriate security measures to safeguard sensitive health information. 

The financial industry is subject to various regulations that aim to protect customer data, prevent fraud, and ensure financial integrity. Regulations such as the Payment Card Industry Data Security Standard (PCI DSS) \cite{kalkan2010payment} and data protection regulations may require robust security measures, including air-gapping in specific scenarios. Industries that operate critical infrastructure, such as energy and utilities, transportation, and water management, often have regulations or standards specific to their sector. While these regulations may not explicitly mention air-gapping, they emphasize the need for robust security controls to protect critical systems and data from cyber threats.

\begin{table}[]
	\caption{Types of high-profile information that might be kept secured, protected, or air-gapped}
	\label{tab:inf}
	\resizebox{\textwidth}{!}{%
		\begin{tabular}{@{}lll@{}}
			\toprule
			\# & Type of high-profile data & Industries/sectors/domains \\ \midrule
			1 & Personally identifiable information (PII) & All \\
			2 & Research and development data & Academy, universities and research centers \\
			3 & Trade secrets and proprietary information & Legal, law, attorney offices, high-tech sector \\
			4 & Intellectual properties (IP) & Legal, law, attorney offices \\
			5 & Medical records, health research and clinical trials & Healthcare, medical centers, hospitals \\
			6 & Finance information (private/organization) & Banks, financial institutions \\
			7 & Cryptographic private keys and certificates & All \\
			8 & Cryptocurrency data, keys, and wallets & Digital currency, cryptocurrency markets, crypto wallets \\
			9 & Governmental information and documents & Governmental and public sectors \\
			10 & Governmental information and documents & Governmental and public sectors \\
			11 & Biometric information & Police, law, public sector \\ \bottomrule
		\end{tabular}%
	}
\end{table}

\subsection{Examples of Private Data Leakage Incidents}
It is known that if data is not kept separate, attackers can breach, compromise, and leak it despite monitoring and DLP solutions. Some notable examples of personal data leaks from recent years are given below.

\textbf{MOVEit Breach (June 2023).} The widespread breach of the file transfer tool, MOVEit, has had a significant impact on over 200 organizations and an estimated 17.5 million individuals as of July 2023 \cite{USenergy22:online}. Several federal agencies, including the Department of Energy, Department of Agriculture, and Department of Health and Human Services, have been affected by this breach. Additionally, it is believed that many schools throughout the United States have also been targeted in this attack. As more details emerge regarding the implications of this incident, additional breaches have been confirmed at Shell, Siemens Energy, Schneider Electric, City National Bank, and several international targets. The attack originated from a security vulnerability within MOVEit's software. Although MOVEit patched the flaw once it was identified, hackers gained unauthorized access to sensitive data. 

\textbf{Ronin Crypto Theft (2022).} In March 2022, it was reported that a blockchain gaming platform that relied on cryptocurrency became the target of a cyberattack \cite{RoninBri85:online}. The platform, known as Ronin's Axie Infinity game, allows players to earn digital currency and non-fungible tokens (NFTs) which are secure financial assets stored on a blockchain. As the game gained popularity, the company lowered security protocols to accommodate a larger player base. While this allowed more players to join, it also provided an opportunity for criminals to exploit the system. They stole a staggering \$625 million worth of cryptocurrency. 

\textbf{LinkedIn Data Breach (2021).} In June 2021, it was reported that a massive data breach had exposed 700 million LinkedIn users \cite{LinkedIn23:online}. The leaked data included names, email addresses, phone numbers, workplace information, and, in some cases, sensitive professional details. This incident raised concerns about user data security on professional networking platforms.

\textbf{Accellion Data Breach (2021).} Accellion, a file-sharing software provider, experienced a data breach in December 2020 and January 2021 \cite{Exploita62:online}. The breach has affected multiple organizations that used Accellion's legacy File Transfer Appliance (FTA), including government agencies, financial institutions, and healthcare organizations. The incident resulted in unauthorized access and theft of sensitive data from affected organizations, such as customer information and intellectual property.

\textbf{T-Mobile Data Breach (2021).} T-Mobile, a major telecommunications company in the United States, announced a data breach in August 2021 \cite{TMobiles9:online}. The breach exposed personal information about approximately 54 million customers, including names, social security numbers, and driver's license information. This incident highlighted the potential risks associated with sensitive customer data exposure in the telecommunications industry.

\textbf{Magellan Health Ransomware Attack (2020).} In April 2020, Magellan Health, a healthcare services provider, fell victim to a ransomware attack \cite{Magellan55:online}. The attackers encrypted the company's systems and exfiltrated sensitive data, including personal and medical information. The incident raised concerns about healthcare organizations' vulnerability to ransomware attacks and the potential impact on patient privacy and healthcare services.

\section{Adversarial Attack Model}
While air-gapped networks provide a significant level of isolation, they are not impervious to attacks. Advanced attackers have demonstrated the ability to utilize different techniques to breach or compromise air-gapped environments. These methods include covert channels, analysis of electromagnetic radiation, and physical infiltration. The attack model targeting air-gapped networks typically consists of three primary phases, which align with the kill chain of Advanced Persistent Threats (APTs): (1) network infiltration, (2) data collection, and (3) data exfiltration.

During the initial stage of the attack, infecting an air-gapped network requires the introduction of malware. There are several approaches that attackers can employ to compromise air-gapped networks. Firstly, physical access represents a significant vulnerability that can be exploited. Attackers can introduce malware into the network using USB drives, CD/DVDs, or other portable devices \cite{dorais2021jumping}\cite{AirGappe49:online}. Secondly, supply chain attacks are another method where the attacker targets the software or hardware supply chain to compromise the network and infect its computers \cite{Beatingt3:online}\cite{Guri2018b}. Configuration errors or human mistakes form a third potential attack vector. For instance, misconfigurations of firewalls or human errors can inadvertently expose network infrastructure to attacks, even within an isolated network. Attackers may also utilize social engineering tactics, including phishing emails or phone calls, to deceive employees and gain access to their credentials, thus introducing malware into the network \cite{Beatingt3:online}. According to media reports, past instances of air-gap breaches such as the cases of Agent.Btz and Stuxnet, involved advanced persistent threats (APTs) that infected air-gapped environments via USB drives.

\section{Air-Gap Exfiltration}

Attackers have various covert channels to exfiltrate keystrokes, documents, images, encryption keys, and biometric information from air-gapped networks. Below we list the techniques and covert channels in this threat domain.

\subsection{Insider Threats Leakage (Physical Access)}
Insider threats are potential risks or harm posed to an organization's security, data, or assets by individuals within it. These individuals may include employees, contractors, or anyone with authorized access to the organization's systems, networks, or sensitive information. Insider threats can arise from various motivations, including financial gain, personal grievances, ideology, or unintentional actions. A few examples reported in the last decade are below. In spite of the fact that the data we kept was secure, the insider may use his credentials to access the data without raising any security alerts.

\begin{itemize}
	\item {\textbf{Edward Snowden leak}}. The Edward Snowden leak refers to the disclosure of classified documents by Edward Snowden, a former contractor for the National Security Agency (NSA) of the United States \cite{bauman2014after}. In June 2013, Snowden leaked a trove of classified information to journalists, exposing widespread surveillance activities conducted by the NSA and its international partners. Snowden's actions had far-reaching implications for cybersecurity, privacy rights, and government transparency. The leaks exposed government surveillance capabilities, raised public awareness about privacy concerns, and triggered significant discussions and reforms in policies and legislation related to surveillance and data collection practices. One method Snowden reportedly used was to access and download classified documents onto portable storage devices.
	
	\item {\textbf{Chelsea Manning leak}}. Manning was a former intelligence analyst in the United States Army who leaked confidential military and diplomatic documents to WikiLeaks in 2010 \cite{chelseaM5:online}. Manning's leaks were significant in terms of impact and classified information volume. She provided WikiLeaks with hundreds of thousands of documents, including classified field reports from the Iraq and Afghanistan wars. She also provided diplomatic cables, and videos showing a U.S. helicopter attack in Iraq.
	
	\item {\textbf{Harold Martin leak}}. Harold Martin, a former contractor for the National Security Agency (NSA) in the United States, was arrested in 2016 for stealing classified documents and data \cite{FormerNS71:online}. The stolen material included highly sensitive national defense and intelligence operations information. Martin's case highlighted the challenges of detecting and preventing insider threats over an extended period.	
	
	\item{\textbf{Paige Thompson leak}}. Paige Thompson, a former employee of a cloud computing company, was charged with hacking into a major financial institution's servers in 2019 \cite{ExAmazon31:online}. Thompson gained unauthorized access to over 100 million customers, exposing sensitive data such as social security numbers and financial records. This incident highlighted the risk of insider threats from individuals with legitimate access to systems and emphasized the importance of robust security controls.
	
\end{itemize}

\subsection{USB Flash Drives}
Malware can use USB devices to jump over air gaps. It can be used to bypass physical or logical barriers implemented to isolate secure or sensitive networks from external connections. Malware can exploit vulnerabilities in the operating system or software handling USB devices. When an infected USB device is connected to a computer within a secure network, malware can automatically execute and infect the system.

\begin{figure}
	
	\centering
	
	\includegraphics[width=\linewidth]{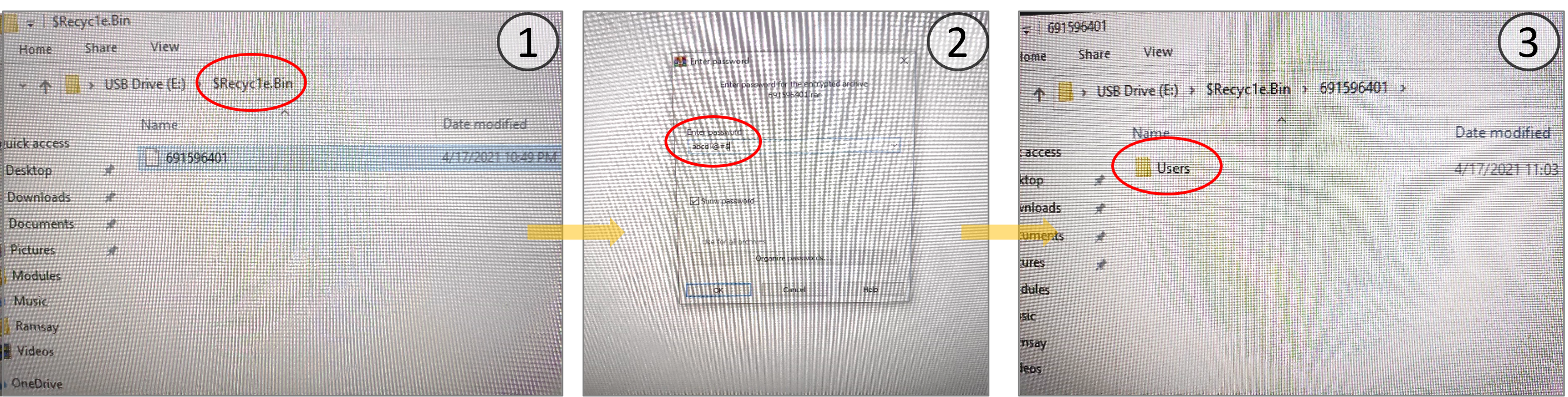}
	\caption{External screenshots of a machine infected with USBCulprit malware. The file is encrypted in \$Recycycle.bin folder.}
	\label{fig:usb}
	
\end{figure}

ESET security company listed more than 17 APTs designed to attack air-gapped networks, including USBStealer, Agent.BTZ \cite{gostev2014agent}, Stuxnet, Fanny, MiniFlame, Flame, Gauss, ProjectSauron, EZCheese, Emotional Simian, USB Thief, USBFerry, Brutal Kangaroo (the CIA leak), Retro, PlugX, and Ramsay \cite{dorais2021jumping}. For example, USBCulprit, discovered in 2020, is an APT observed using removable devices, specifically USB flash drives \cite{guri2021usbculprit}\cite{Sednitus4:online}. It uses these devices to reach air-gapped systems and exfiltrate information with its extensive tool set. The malware can scan multiple paths on an infected system, look for document and presentation files (.docx, .pptx, .pdf, etc.), and hide them encrypted on a removable USB drive (Figure \ref{fig:usb}). This allows it to move laterally to other systems when said USB drive is plugged into another network. 

Agent.btz is another malware that infects U.S. classified military networks and spreads through removable USB drives and other external storage devices \cite{gostev2014agent}. When an infected USB drive is inserted into a computer, malware can automatically execute and spread to other connected systems. It exploits vulnerabilities in the Windows operating system to spread across networks. 

Brutal Kangaroo is another hacking tool and framework that was part of the Vault 7 leaks \cite{ramirezdiscovering}. Brutal Kangaroo was designed as an attack methodology to target air-gapped networks, which are isolated from the internet or other external networks. It aims to bridge the gap between a compromised internet-connected computer (the "sender") and the target air-gapped network (the "receiver"). However, due to its security holes, USB drives might be limited in certain environments. For example, due to Agent.Btz incident in 2008, the DOD banned removable, flash-type drives on all government computers. 

In June 2023, CheckPoint reported a malware variant, WispRider, which propagates through USB and enables infiltration of potentially isolated systems \cite{Stealthy95:online}.

\subsection{Acoustic Attacks}
Acoustic covert channels are communication channels that allow information to be transmitted covertly or hidden using sound or audio signals. In cybersecurity and data leakage, sound-based covert channels send data through audio frequencies imperceptible to human ears. In most acoustic covert channels, attackers use ultrasonic frequencies above human hearing limits to transmit data. 

Several works in the past have shown how attackers can exploit ultrasonic waves to exfiltrate sensitive information from air-gapped systems, including keylogging and documents. In these attacks, the malware uses computer audio hardware and loudspeakers to encode information on top of sonic and ultrasonic signals, transmitting it between computers \cite{hanspach2014covert} and smartphones \cite{Deshotels2014}. 

However, in some systems, audio hardware might be disabled to create audio-gap protection \cite{AirGapCo76:online}\cite{Jumpingt83:online}. Researchers show how attackers can control CPU \cite{guri2020fansmitter} and GPU \cite{guri2022gpu} fans' speed to generate acoustic noise to overcome this limitation. They encode information in fan rotations per minute (RPM) noise. A nearby smartphone can receive the information and decode the transmitted data. Similarly, attackers use the noise generated by the Hard-Disk Drive (HDD) \cite{guri2017acoustic} and the CD/DVD drives \cite{guri2020cd} to encode sensitive information from audio-less computers. In 2022 Briseno et al. proposed using inkjet printers to exfiltrate arbitrary sensitive data by producing mechanical acoustic signals \cite{de2022inkfiltration}. Distances of up to 4 meters allowed an average low bit rate. MOSQUITO attack \cite{Guri2018Mosquito} enables the transmission of data between two air-gapped computers through ultrasonic sound waves emitted by the speakers and picked up by the manipulated speaker of the receiving computer. Despite the absence of microphones, the authors maintain a covert communication channel via a speaker-to-speaker protocol \cite{guri2020speaker}. In 2021, Guri presented an attack named POWER-SUPPLAY, in which malware in the compromised computer can exploit the power supplies and use them as an out-of-band speaker to transmit ultrasound and leak information \cite{guri2021power}. Workstation capacitors generate acoustic noise and encode binary and textual data in this case.

\subsection{Electromagnetic and Electric Attacks}
\begin{figure}
	\centering
	\includegraphics[width=0.6\linewidth]{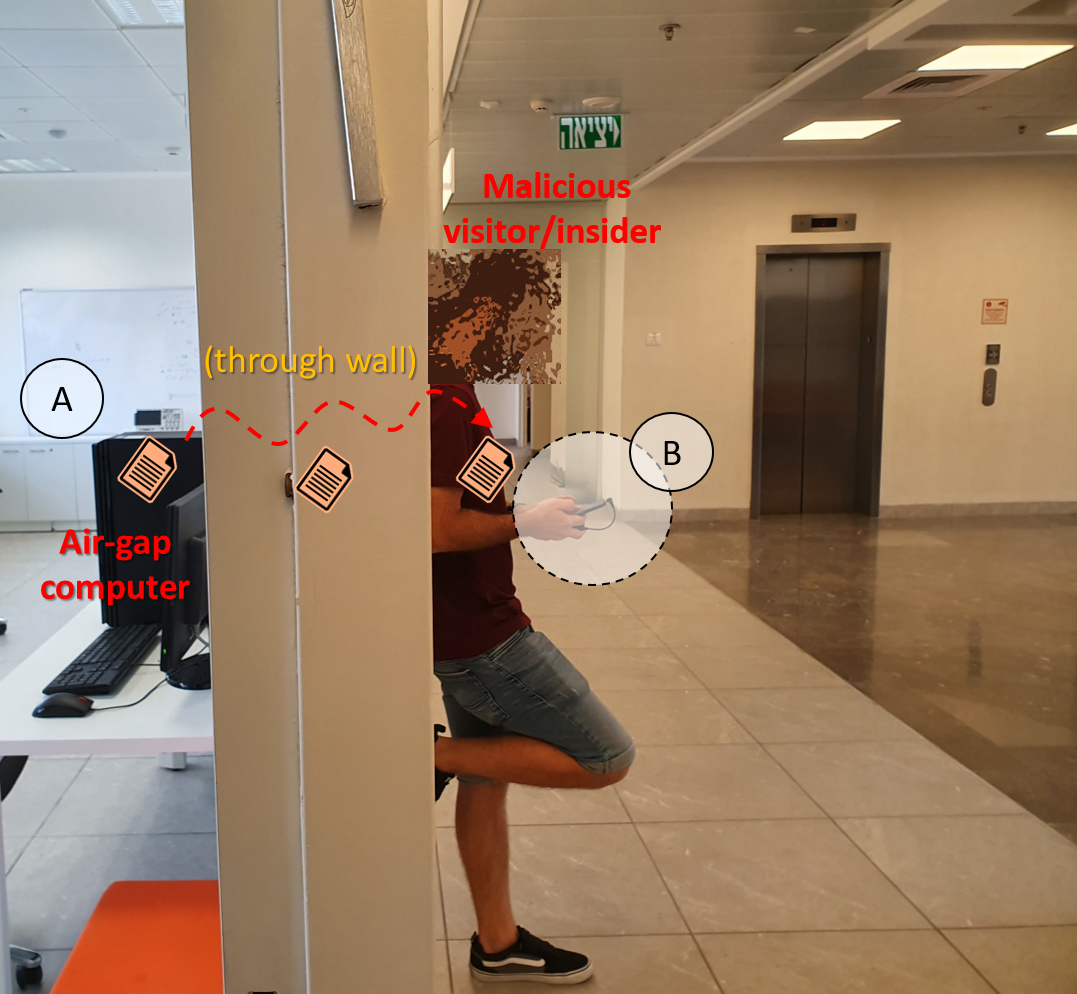}
	\caption{Attack scenario: An APT running on a contaminated air-gapped computer (A) exfiltrates a sensitive file via electromagnetic signals. The file is received by the smartphone of a malicious insider/visitor located behind the wall (B).}
	\label{fig:scenario}
\end{figure}

Electromagnetic covert channels refer to a method of communication that exploits electromagnetic radiation from various system parts to transmit information from air-gapped computers (Figure \ref{fig:scenario}). These channels have been extensively studied for many years. In a seminal work, Kuhn introduced video cards to broadcast radio signals modulated with information from workstations \cite{kuhn1998soft}. Specifically in the context of air-gap attacks, the AirHopper attack leveraged video cards in air-gapped computers to generate FM signals carrying leaked information \cite{Guri2014}. Similarly, other notable attacks such as GSMem \cite{Guri2015}, USBee \cite{Guri2016}, BitJabber \cite{zhan2020bitjabber} EMLoRa \cite{shen2021lora}, and Air-Fi \cite{guri2022air} introduced attack scenarios in which attackers utilized different radiation sources from air-gapped systems for data exfiltration. LANTENNA, presented in 2021, allows data exfiltration from air-gapped networks via the emission of Ethernet cables \cite{guri2021lantenna}. In 2022, researchers presented SATAn, an attack that uses the SATA cable of HDD/SSD drives as an antenna to exfiltrate information from air-gapped computers \cite{guri2022satan}. Additionally, electricity can be another source of data leakage. In 2019, researchers introduced "PowerHammer," a method that enables data exfiltration through conducted emissions to power lines \cite{guri2019powerhammer}. Furthermore, in 2020, Shao et al. demonstrated the use of noise-induced power lines for data transmission \cite{shao2020your}.

\subsection{Magnetic Attacks}
Faraday cages can mitigate the risk of electromagnetic radiation. Researchers have proposed utilizing magnetic fields for covert communication to overcome this, enabling data leakage from Faraday-caged air-gapped computers to nearby magnetic sensors. ODINI \cite{guri2019odini} and MAGNETO \cite{guri2018magneto} are two types of attacks that allow the exfiltration of data from air-gapped, Faraday-shielded workstations and servers. These attacks use near-field magnetic emission from the CPU cores and modulate binary information. Nearby smartphones or magnetic sensors can receive magnetic signals. Matyunin presented a magnetic covert channel that uses the emission induced by different hard-disk drives (HDD) during I/O operations such as reading and writing \cite{matyunin2016covert}. MagView \cite{zhang2020magview} is another magnetic covert channel that uses the emanation from screen videos to encode data covertly.

\subsection{Optical Attacks}
\begin{figure}
	\centering
	\includegraphics[width=0.6\linewidth]{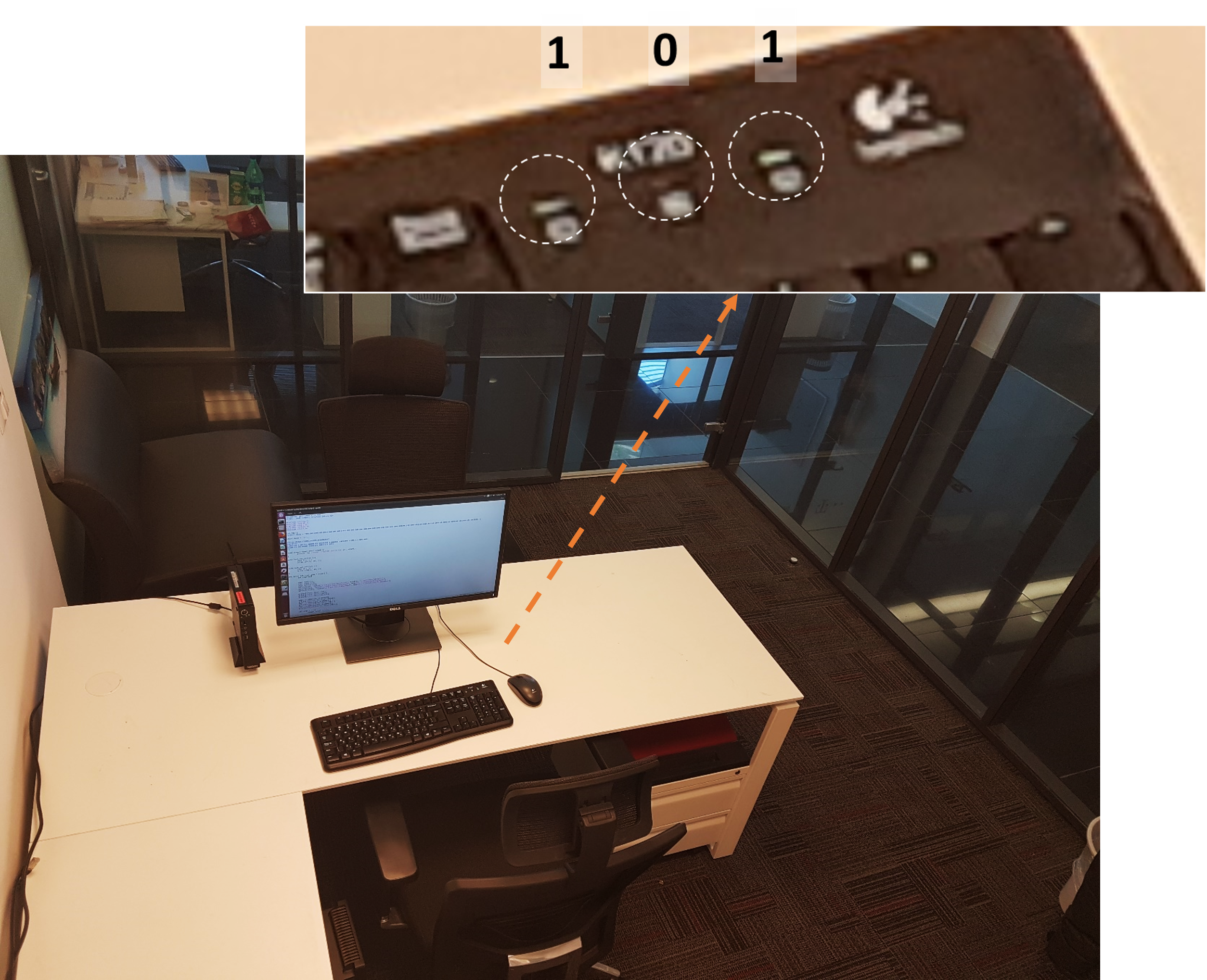}
	\caption{Optical covert channel via keyboard LEDs.}
	\label{fig:key}
\end{figure}

Optical covert channels are widely studied communication channels that exploit light or optical signals to transmit information covertly. These channels utilize various techniques to encode and transmit data using light as the carrier medium. Information can be incorporated into light signals by modulating their intensity, frequency, or phase. For instance, the intensity of light emitted by an LED or the flickering patterns of a display screen can be manipulated to send binary data. Research shows that data can be encoded in optical signals covertly sent from keyboard status LEDs \cite{guri2019ctrl} (Figure \ref{fig:key}). The speed can reach hundreds of bits/sec if an appropriate receiver is employed (e.g., an optical sensor). Other types of attacks use routers \cite{guri2018xled}, network cards (NIC), and HDD status LEDs \cite{Guri2017} to modulate binary and textual information. Remote cameras can receive data up to tens or hundreds of meters away via drone cameras. AIR-Jumper is a particular air-gap covert channel that uses the infrared light in a security camera to generate bidirectional data exchange between camera devices over the air-gap \cite{guri2019air}. Computer screens can also be used to modulate information through the air-gap via invisible images or slight changes in brightness \cite{guri2019brightness}.

\subsection{Thermal Attacks}
Thermal covert channels are a type of covert communication channel that exploits variations in temperature or thermal emissions to transmit data hidden or covertly. These channels take advantage of the fact that electronic devices generate heat during their normal operations, which can be manipulated to encode and transmit information.
In BitWhispr \cite{guri2015bitwhisper}, thermal-based communication is maintained between two adjacent air-gap computers. By using the smartphone's temperature measurement capabilities, HotSpot \cite{guri2019hotspot} received the information.

\subsection{Vibration Attacks}
Vibration covert channels are a type of covert communication channel that exploits variations in mechanical vibrations to transmit data in a hidden or covert manner. These channels take advantage of the fact that electronic devices and their components generate vibrations during their normal operations, and these vibrations can be manipulated to encode and transmit information. In 2021, Guri presented Air-Viber \cite{guri2021exfiltrating}, an attack that uses table vibration to exfiltrate information from air-gapped workstations. The data can be received via an accelerator installed on any standard smartphone. GAIROSCOPE attack \cite{guri2021gairoscope} measures the smartphone's gyroscope sensor's small vibrations at resonance frequency to create an ultrasonic vibration-based air-gap attack.

\section{Protection and Mitigation}
Countermeasures against air-gap covert channels aim to prevent or mitigate data exfiltration or unauthorized communication from air-gapped systems. While eliminating covert channels is challenging, combining technical and procedural measures can significantly enhance security. Countermeasures can be categorized into regulatory and technological.

\textbf{Physical isolation.} Maintain strict physical access controls to secure sensitive areas and restrict unauthorized individuals from physically tampering with or connecting unauthorized devices to air-gapped systems. Physical measures may include shielding, Faraday cages, soundproofing, temperature control, and vibration dampening to reduce the risk of unintended emissions or interceptions.

\textbf{Red-black separation.} Red-black separation, also known as red-black isolation or red-black architecture, is a security principle and design approach used to mitigate the risk of data leakage or unauthorized communication between classified or sensitive systems and non-classified or less secure systems. The term "red" refers to systems that are classified or sensitive. In contrast, "black" refers to non-classified or less secure systems. The goal of red-black separation is to enforce strict isolation between the two types of systems to prevent the unauthorized transfer of information or the compromise of sensitive data. This is particularly crucial in environments where sensitive information needs to be protected from potential threats, such as military or government organizations.

\textbf{Device hardening.} In the device hardening approach, the defender configures and harden devices in air-gapped systems to minimize unintentional emissions, turn off unnecessary ports or interfaces, and closely monitor and control connections to external devices. For example, it prohibits Wi-Fi or wireless in certain environments. It's important to note that device hardening may cause a degradation of the device's capabilities, e.g., an audio-gapped device will not be able to play sound. This makes the solution less practical on a wide scale.

\textbf{Signal Monitoring and Detection.} Deploy specialized monitoring tools and technologies to detect unusual electromagnetic, acoustic, thermal, or vibration emissions that may indicate the presence of covert channels. This includes using intrusion detection systems (IDS), network traffic analysis, anomaly detection algorithms, and physical sensors. However, previous work showed that signal monitoring is prone to high levels of false-positives due to environmental noise. For example, other computers and screens emanate unintentional electromagnetic emissions.

\textbf{Operating System Behavioral Analysis.} It is possible to implement behavior-based monitoring and analysis techniques to detect unusual or unauthorized activities within air-gapped systems, such as unexpected data transfers or unique resource utilization patterns. For example, to mitigate the ultrasonic attack, it has been proposed to implement a so-called 'ultrasonic firewall' to block the ultrasonic sound. However, these detection methods have a fundamental weakness since they can be bypassed by malware with high privileges.

\textbf{Employee Education and Awareness.} Provide regular training and awareness programs to employees to educate them about the risks of covert channels. In addition, educate them about the importance of adhering to security protocols, and detecting and reporting suspicious activities. It is also essential to conduct regular security assessments, including penetration testing, to identify vulnerabilities and evaluate countermeasure effectiveness. This helps ensure ongoing security and adaptation to emerging threats on air-gapped and secure networks.

\section{Conclusion}
Private information refers to sensitive personal, confidential, or restricted information about individuals or organizations. This includes Personally Identifiable Information (PII), health information, financial data, and legal documents, among others.
To keep this data secure from online threats, an organization may keep it in a so-called `air-gapped' network. Air gap refers to a physical and logical separation from the Internet and other networks. We show that a capable, motivated adversary can breach the air-gapped network despite the isolation. Although there is no network connectivity, attackers can use various special air-gap covert channels to exfiltrate data. This includes using acoustic, electromagnetic, electric, optical, thermal, and physical mediums to encode data and leak it secretly. We review previous in-wild malware capable of infecting air-gapped networks. We present the adversarial attack model on air-gap networks, categorize the different covert channels, and describe countermeasures. Despite the high level of isolation, the information in air-gapped networks is not hermetically protected from air-gap covert channels. Additional defensive measures should be taken to protect data from cyber threats.

\bibliographystyle{plain}

\end{document}